\documentclass[aps,prl,twocolumn,showpacs,preprintnumbers,amsmath,amssymb]{revtex4-1}
 \def\ep{{\epsilon}}

 \def\frac#1#2{{#1\over #2}}

 \def\s{\sqrt}

\def\be{\begin{equation}}
\def\ee{\end{equation}}
\def\ba{\begin{eqnarray}}
\def\ea{\end{eqnarray}}

 \def\de{\partial}

 \def\ti{\tilde}

 \def\la{\langle}
 \def\lb{\rangle}
 \def\ep{\epsilon}

\usepackage{color}
\usepackage{graphicx}% Include figure files
\usepackage{dcolumn}% Align table columns on decimal point
\usepackage{bm}% bold math

\begin{document}

\title{Entanglement entropy in holographic moving mirror and Page curve}
\preprint{YITP-20-149; IPMU20-0121}
\author{Ibrahim Akal$^{a}$, Yuya Kusuki$^{a}$,
Noburo Shiba$^{a}$, Tadashi Takayanagi$^{a,b,c}$ and Zixia Wei$^{a}$}
\affiliation{$^a$Yukawa Institute for Theoretical Physics, Kyoto University, \\
Kitashirakawa Oiwakecho, Sakyo-ku, Kyoto 606-8502, Japan}
\affiliation{$^b$Inamori Research Institute for Science, 620 Suiginya-cho, Shimogyo-ku, Kyoto 600-8411 Japan}
\affiliation{$^{c}$Kavli Institute for the Physics and Mathematics of the Universe,\\ University of Tokyo, Kashiwa, Chiba 277-8582, Japan}
%\date{\today}
\begin{abstract}
We calculate the time evolution of entanglement entropy in two dimensional conformal field theory with a moving mirror. For a setup modeling Hawking radiation, we obtain a linear growth of entanglement entropy and show that this can be interpreted as the production of entangled pairs. For a setup, which mimics black hole formation and evaporation, we find that the evolution follows the ideal Page curve. We perform these computations by constructing the gravity dual of the moving mirror model via holography. We also argue that our holographic setup provides a concrete model to derive the Page curve for black hole radiation in the strong coupling regime of gravity.
\end{abstract}

%\pacs{72.10.-d,73.21.-b,73.50.Fq}
\maketitle
%%%%%%%%%%%%%%%%%%%%%%%%%%%%%
\section{Introduction}
%%%%%%%%%%%%%%%%%%%%%%%%%%%%%

Moving mirrors have been known for a while as a class of instructive models that mimic Hawking radiation \cite{Hawking:1974sw} based on quantum field theory \cite{Birrell:1982ix,Davies:1976hi} where unitarity is manifest. On the other hand, in the case of black hole evaporation, it has been a significant problem to understand whether unitarity is maintained  
in the gravitational theory. One manifestation of unitary black hole evaporation is the Page curve for the entropy of Hawking radiation \cite{Page:1993df}. Based on the fine grained entropy formula \cite{RT,HRT,Faulkner:2013ana,Engelhardt:2014gca}, this has been derived semiclassically for field theories coupled to gravity \cite{Penington:2019npb,Almheiri:2019psf,Almheiri:2019hni} and confirmed by direct gravity replica computations \cite{Penington:2019kki,Almheiri:2019qdq}. See e.g. \cite{Akers:2019nfi,Almheiri:2019yqk,Rozali:2019day,Chen:2019uhq,Bousso:2019ykv,Kusuki:2019hcg,Pollack:2020gfa,
Liu:2020gnp,Marolf:2020xie,Piroli:2020dlx,Balasubramanian:2020hfs,Verlinde:2020upt,
Chen:2020wiq,Gautason:2020tmk,Anegawa:2020ezn,Giddings:2020yes,Hashimoto:2020cas,
Hartman:2020swn,Agon:2020fqs,Hollowood:2020cou,Krishnan:2020oun,Alishahiha:2020qza,
Geng:2020qvw,Chen:2020uac,Almheiri:2020cfm,Li:2020ceg,Chandrasekaran:2020qtn,Bak:2020enw,Bousso:2020kmy,
Dong:2020uxp,Akal:2020wfl,Engelhardt:2020qpv,Karlsson:2020uga,Chen:2020jvn,Chen:2020tes,
Hartman:2020khs,Murdia:2020iac,Altland:2020ccq,Balasubramanian:2020xqf,Sybesma:2020fxg,Stanford:2020wkf,
Ling:2020laa,Chakravarty:2020wdm,Bhattacharya:2020uun,Harlow:2020bee,Chen:2020ojn,Kirklin:2020zic,Goto:2020wnk,Hsin:2020mfa} for further progress along this direction. For recent related works refer to \cite{Nomura:2019qps,Laddha:2020kvp,Akal:2020ujg}.

In this article, we first present concrete calculations of entanglement entropy in moving mirror setups and show that this leads to an ideal Page curve. This itself provides a novel nonequilibrium setup, where quantum entanglement evolves rapidly. Moreover, we present a close connection between moving mirror models and black hole radiation via a particular version of the anti de-Sitter (AdS)/conformal field theory (CFT) correspondence \cite{Ma}, namely, in the case when the CFT is defined on a manifold with a boundary \cite{Karch:2000gx,Ta}. For earlier studies of entanglement entropy in moving mirror models refer to \cite{Bianchi:2014qua,Hotta:2015huj,Good:2016atu,Chen:2017lum,Good:2019tnf}.

\begin{figure}[h]
  \centering
  \includegraphics[width=0.28\textwidth]{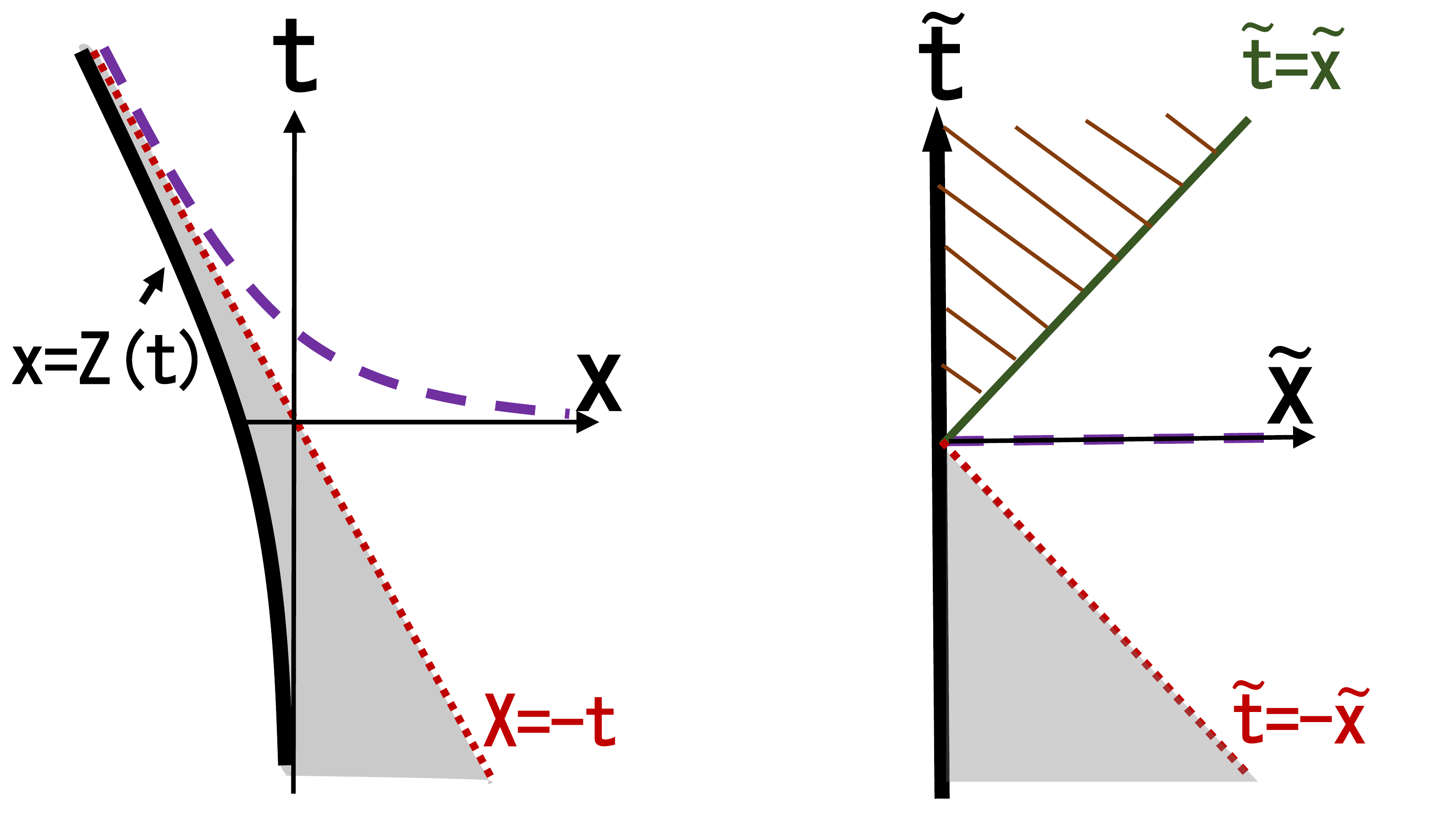}
  \caption{The moving mirror setup (left) and its conformal transformation into a static mirror (right). 
The Mirror trajectory is depicted by the thick curve. The shaded region shown in the right panel 
corresponds to an inside horizon region, which is missing in the left picture.}
\label{mvsetupfig}
\end{figure}

%%%%%%%%%%%%%%%%%%%%%%%%%%%%%
\section{Moving mirror from conformal maps}
%%%%%%%%%%%%%%%%%%%%%%%%%%%%%

A moving mirror setup in two dimensions is specified by the trajectory of a mirror profile $x=Z(t)$. 
We consider a CFT which lives on the right region, i.e. $x\geq Z(t)$. 
A conformal transformation (here, we set $u=t-x$ and $v=t+x$) \cite{Birrell:1982ix,Davies:1976hi},
\ba
\ti{u}=p(u),\ \ \ \ti{v}=v,  
\label{cfmap}
\ea 
maps this into a simple setup with a static mirror  $\ti{u}-\ti{v}=0$, as depicted in Fig.~\ref{mvsetupfig}.
Here, we choose the function $p(u)$ such that the mirror trajectory is given by $v=p(u)$, i.e.
\ba
t+Z(t)=p\left(t-Z(t)\right).
\ea
For example, we can calculate the energy stress tensor from the 
conformal anomaly via the map \eqref{cfmap}, such that
\ba
T_{uu}
=
\frac{c}{24 \pi} \left( \frac{3}{2} \left(  \frac{p''(u)}{p'(u)} \right)^2 - \frac{p'''(u)}{p'(u)}  \right),
\label{EMT}
\ea
where the components $T_{uv}$ and $T_{vv}$ are vanishing.

As an example of a CFT, consider a massless free scalar $\phi$. We impose the Dirichlet boundary condition $\phi(t,Z(t))=0$ along the mirror trajectory. A complete set of positive frequency solutions to the equations of motion $\de_u \de_v\phi=0$, which satisfy the latter boundary condition, reads
\begin{equation}
\begin{split}
\phi_\omega(t, x)=i(4\pi \omega)^{-1/2}\left(e^{-i\omega v}-e^{-i\omega p(u)}\right).
\end{split}  \label{complete set}
\end{equation}
Then,  $\phi$ can be expanded in terms of these modes \cite{Birrell:1982ix} as
\begin{equation}
\begin{split}
\phi(t, x)=\int_0^{\infty} d\omega \left[a_\omega^\text{in}\phi_\omega+a_\omega^{\text{in} \dagger}\phi_\omega^{*} \right],
\end{split}  \label{mode expansion}
\end{equation}
where $a_\omega^\text{in}$ and $a_\omega^{\text{in} \dagger}$ are the annihilation and creation operators, respectively.
The in-coming vacuum $|0_\text{in}\lb$ is defined by the  state annihilated by $a_\omega^\text{in}$ and the out-going
vacuum $|0_\text{out}\lb$ is given by a Bogoliubov transformation of  $|0_\text{in}\lb$. The expectation value of the energy stress tensor, i.e. $\la 0_\text{in}|T_{uu}|0_\text{in}\lb$, reproduces \eqref{EMT} for $c=1$.

\begin{figure}[h]
  \centering
  \includegraphics[width=0.2\textwidth]{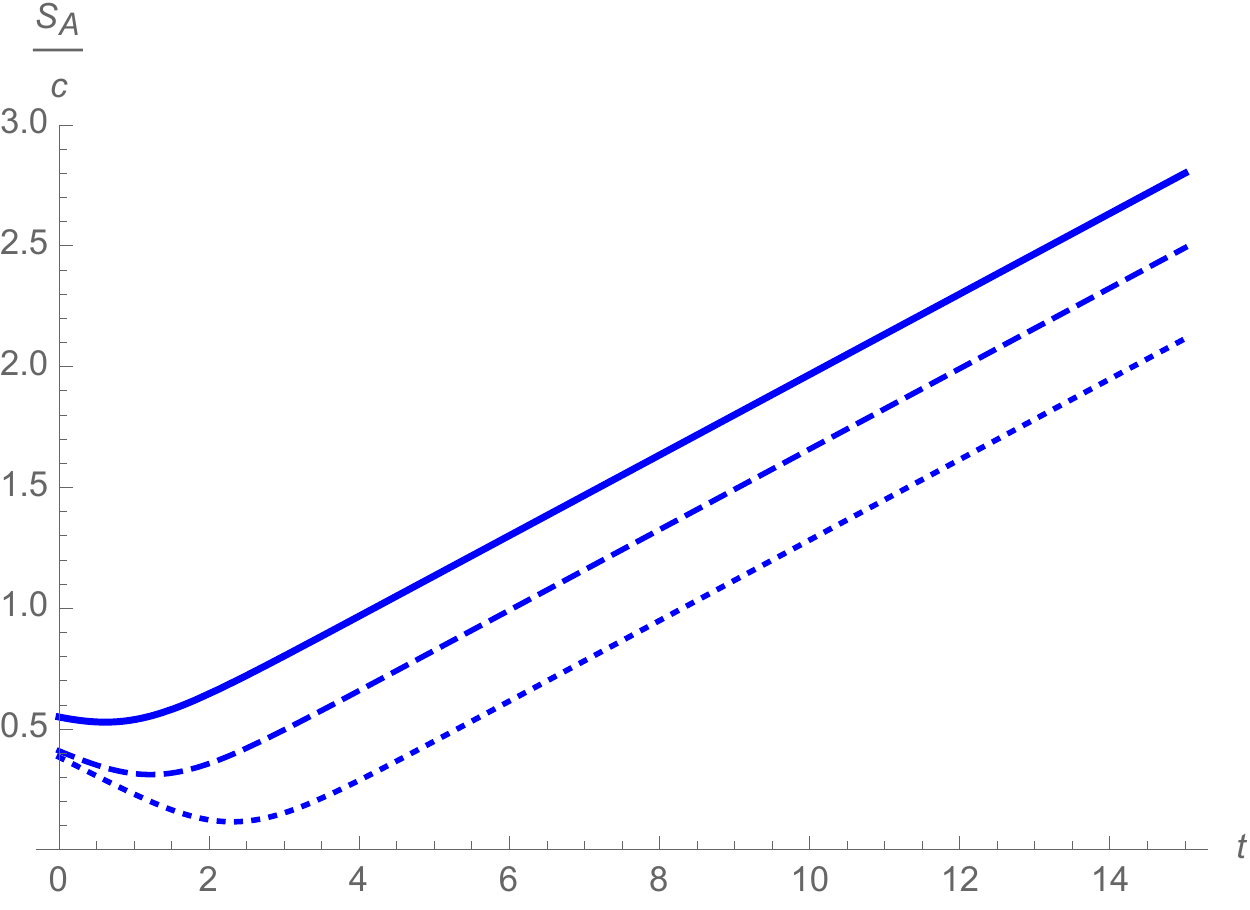}\qquad
  \includegraphics[width=0.16\textwidth]{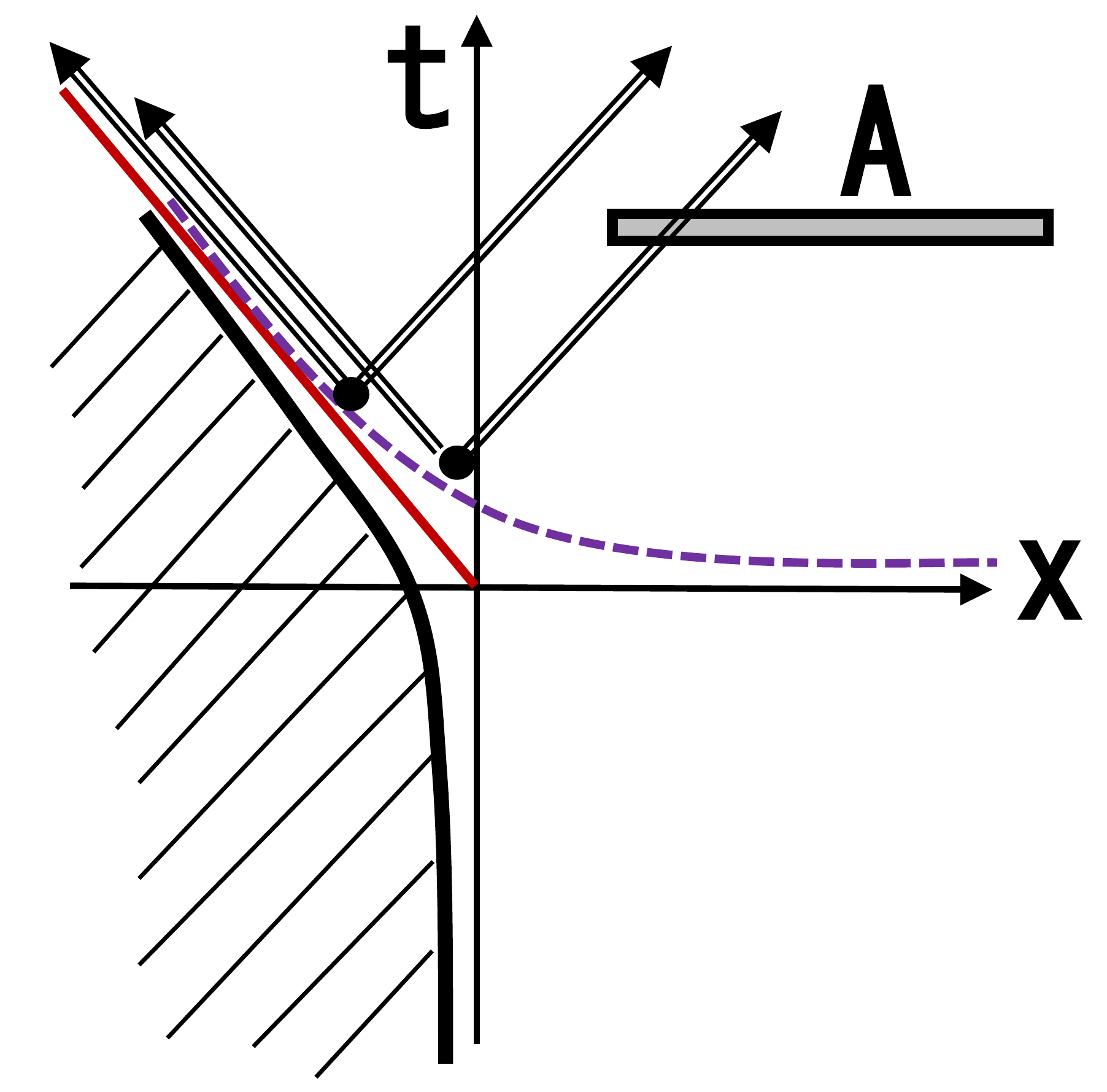}
  \caption{The graphs in the left figure show the time evolution of entanglement entropy $S_A$.
We choose the end point of $A$ to be $x_0=-t+\xi_0$, with $\xi_0=1$ (thick), 
$\xi_0=0.1$ (dashed) and $\xi_0=0.01$ (dotted). We set $\beta=1$, $\ep=0.1$ and $S_\text{bdy} = 0$.
The right figure shows a quasi-particle picture of entanglement growth for the moving mirror. 
The black thick curve represents the mirror trajectory $x=Z(t)$. 
The purple dotted curve describes a spacelike curve defined by $v+p(u)=0$. The red line 
corresponds to the null line. The entangled pair production occurs on the purple dashed curve.}
\label{eeffig}
\end{figure}

%%%%%%%%%%%%%%%%%%%%%%%%%%%%%
\section{Moving mirror with horizon}
%%%%%%%%%%%%%%%%%%%%%%%%%%%%%

For a typical example which models Hawking radiation from a black hole, we choose
\ba
p(u)=-\beta\log(1+e^{-\frac{u}{\beta}}),  
\label{traj}
\ea
where the parameter $\beta$ plays the role of an inverse temperature. Its profile is depicted in the left 
picture of Fig.~\ref{mvsetupfig}. In the early time limit $t\to -\infty$, we have $Z(t)\simeq 0$, while 
in the late time limit, the mirror trajectory gets almost lightlike, $Z(t)\simeq -t-\beta e^{-2t/\beta}$.
As depicted in Fig.~\ref{mvsetupfig}, the region $\ti{u}\geq 0$ in the extended coordinates
is missing for the original coordinates. This region is analogous to the 
inside horizon region in the black hole formation process.

The energy stress tensor \eqref{EMT} reads
\ba
T_{uu}=\frac{c}{48\pi\beta^2}\left(1-\frac{1}{(1+e^{u/\beta})^2}\right),
\label{EMTm}
\ea
which vanishes at early time $u\to -\infty$, and becomes a constant thermal flux, $T_{uu}\simeq \frac{c}{48\pi\beta^2}$, at late time $u\to\infty$.

Let us calculate the entanglement entropy $S_A$ for a semi infinite subsystem $A$ given by  $[x_0,\infty]$ at time $t$. We can calculate $S_A$ from the one point function of the twist operator \cite{CC04,HLW,Casini:2009sr} on the upper half plane $\ti{x}>0$ by using the conformal map \eqref{cfmap} via the replica method. We find
\ba
S_A=\frac{c}{6}\log\frac{t+x_0-p(t-x_0)}{\ep\s{p'(t-x_0)}}+S_\text{bdy}.
\label{sahee}
\ea
Here, $\ep$ is the UV cutoff (lattice spacing) of the CFT and $S_\text{bdy}$ is the boundary entropy \cite{Affleck:1991tk,CC04,Ta}. Note that this formula holds for any CFT.

If we fix the end point of the subsystem $A$, i.e. $x_0$,  
we can approximate \eqref{sahee} at late time $t\to\infty$, and find
\ba
S_A\simeq \frac{c}{12\beta}(t-x_0)+\frac{c}{6}\log\frac{t}{\ep}+S_\text{bdy}.  
\label{apphee}
\ea
The first term linear in $t$ arises from entangled pair production due to the moving mirror, while
the second $\log t$ term comes from the standard vacuum entanglement as the length 
of the complement of $A$ grows linearly.

To study the first contribution in more detail, we allow changing the value of $x_0$ time dependently as
\ba
x_0(t)=-t+\xi_0.  
\label{dfferessnt}
\ea
In the late time limit $u\to \infty$, we obtain
\ba
S_A =\!\frac{c}{6}\!\log\!\frac{\left(\xi_0\!+\!\beta e^{-(t\!-\!x_0(t))/\beta}\!\right)\s{1+e^{(t\!-\!x_0(t))/\beta}}}{\ep}
+S_\text{bdy}.\notag \\
\ea
We choose $\xi_0$ to be positive, but sufficiently small. 
If the left end point of $A$, given by $(u,v)=(2t-\xi_0,\xi_0)$, satisfies $v+p(u)>0$, 
we get the linear growth (see the left panel in Fig.~\ref{eeffig}),
\ba
S_A\simeq \frac{c}{6\beta}t+\frac{c}{6}\log\frac{\xi_0}{\ep}+S_\text{bdy}.
\label{apphpee}
\ea
In this way, we may conclude that the entangled pair production occurs along the spacelike curve 
$v+p(u)=0$, and the propagation of the entangled pairs gives the linear growth of the entanglement entropy 
\eqref{apphpee}. This is sketched in the right panel of Fig.~\ref{eeffig}.
We can also confirm this from the free scalar example \eqref{mode expansion}, where the spacial distribution of the pair production looks like
\begin{align}
&\la 0_\text{in}|\phi(u_1,v_1)\phi(u_2,v_2) \int d\omega a_\omega^{\text{in} \dagger} a_\omega^{\text{in} \dagger}|0_\text{in}\lb
\!\propto\\
&\int\!\frac{d\omega}{\omega}
\!\left[\!e^{\!-i\omega(\!v_1\!+\!p_2\!)}\!+\!e^{\!-i\omega(\!v_2\!+\!p_1\!)}
\!-\!e^{\!-i\omega(\!v_1\!+\!v_2\!)}\!-\!e^{\!-i\omega(\!p_1\!+\!p_2\!)}\!\right], \notag
\end{align}
where we have defined $p_i := p(u_i)$ for brevity.
The first two terms are divergent at $v_1+p_2=0$ and $v_2+p_1=0$. This shows that the entangled pairs are produced along the curve $v+p(u)=0$, and they propagate in opposite directions at the speed of light.

%%%%%%%%%%%%%%%%%%%%%%%%%%%%%
\section{AdS/BCFT and entanglement entropy}
%%%%%%%%%%%%%%%%%%%%%%%%%%%%%

To compute $S_A$ for generic subsystems, we need to specify the target CFT.  
For holographic CFTs, we can calculate $S_A$ via the gravity dual of a CFT defined on a manifold $M$
with a boundary $\de M$, i.e. boundary CFT (BCFT) \cite{Cardy2004}, known as AdS/BCFT \cite{Ta}. In this description,  
the dual geometry is given by extending the boundary $\de M$ into the bulk AdS, 
which leads to a codimension one surface $Q$, called the end of the world brane. This surface $Q$ obeys 
the Neumann boundary condition 
\ba
K_{ab}-h_{ab}K+{\cal T}h_{ab}=0, 
\label{NBC}
\ea
where $h_{ab}$ is the induced metric and $K_{ab}$ is the extrinsic curvature. 
The parameter ${\cal T}$ is the tension of the
brane $Q$ and depends on the boundary condition of the CFT at $\de M$. 
The condition \eqref{NBC} implies the presence of boundary conformal invariance.
Refer to \cite{Takayanagi:2020njm} for an equivalent formulation using 
Chern-Simons gravity, and to \cite{Sully:2020pza} for comparisons with CFT calculations.

\begin{figure}[h]
  \centering
  \includegraphics[width=0.2\textwidth]{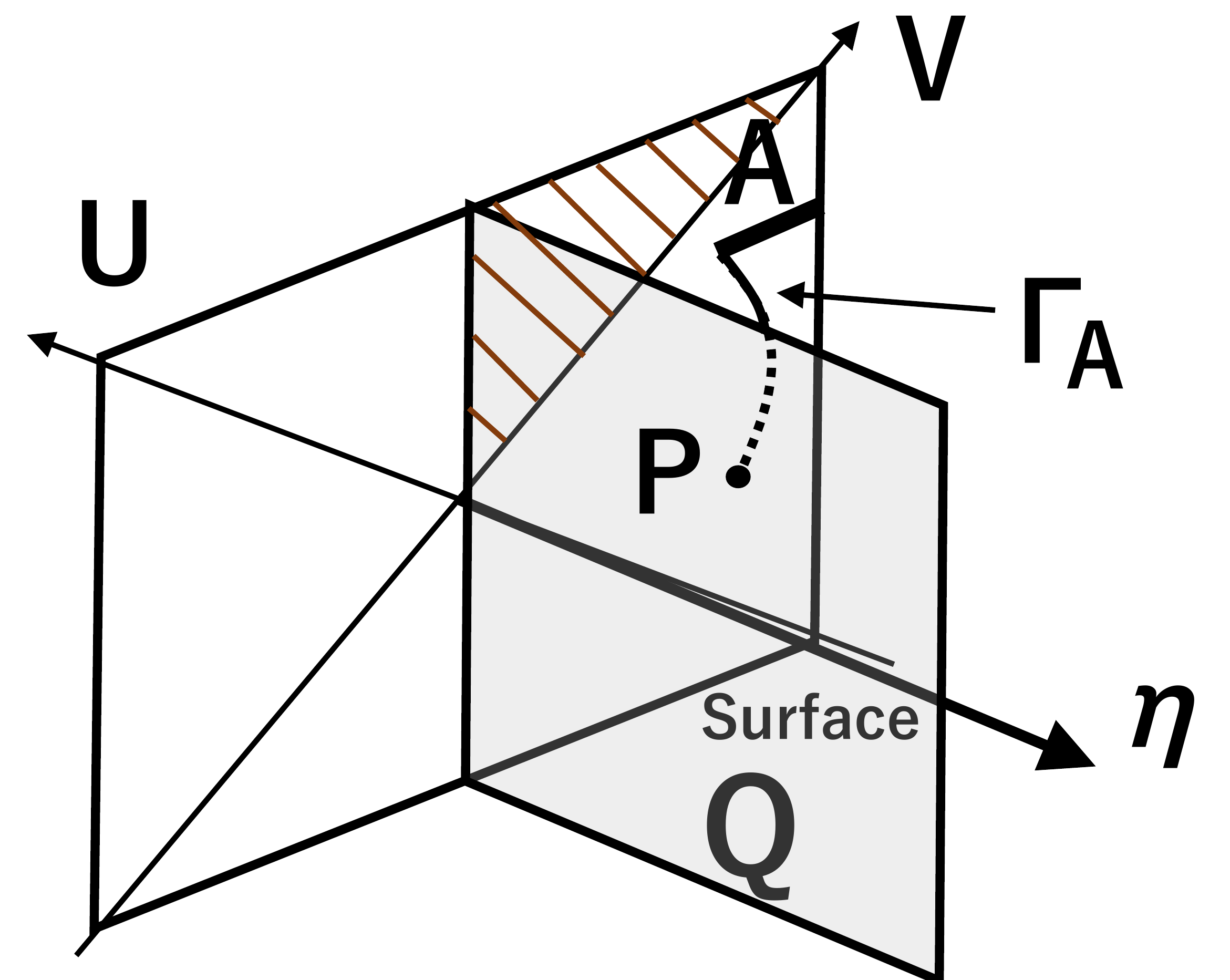}\qquad
   \includegraphics[width=0.2\textwidth]{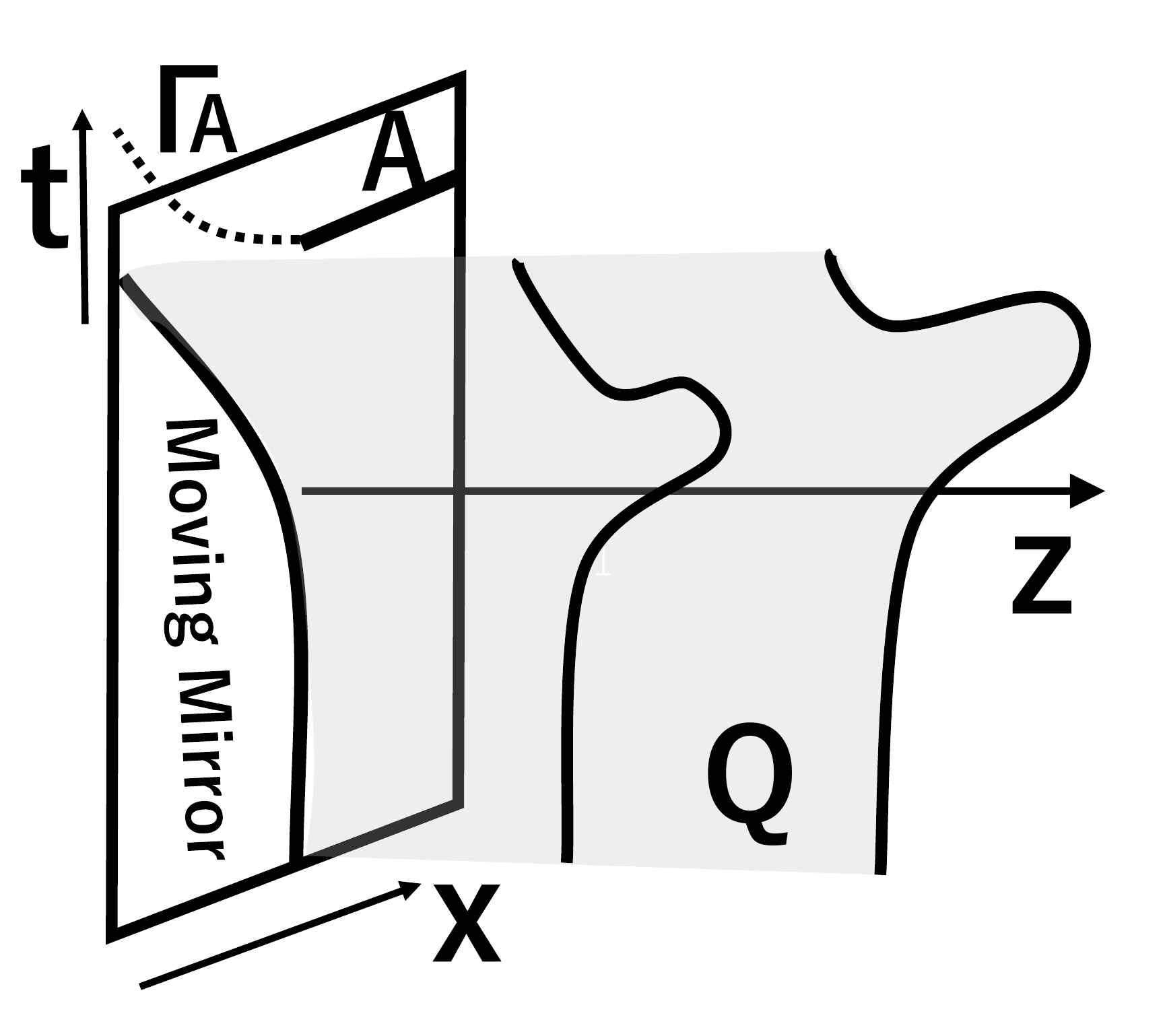}
  \caption{Gravity dual of a moving mirror in the coordinates $(U,V,\eta)$ (left) and $(u,v,z)$ (right). 
We set ${\cal T}=0$. We also show the computation of holographic entanglement entropy.}
\label{holsetupfig}
\end{figure}

We can find a gravity dual by applying 
the following coordinate transformation, which is a special case of \cite{Banados:1998gg,Roberts:2012aq}
and is a bulk extension of the map \eqref{cfmap},
\ba
U=p(u),\quad V=v+\frac{p''(u)}{2p'(u)}z^2,\quad \eta=z \s{p'(u)} 
\label{cordch}
\ea
on Poincare AdS$_3$
\ba
ds^2=\frac{d\eta^2-dUdV}{\eta^2}.  
\label{pol}
\ea
Using \eqref{EMT}, this leads to the metric
\ba
ds^2=\frac{dz^2}{z^2}+\frac{12\pi}{c}T_{uu}(u)(du)^2-\frac{1}{z^2}dudv.
\label{metkads}
\ea
In Poincare AdS$_3$ \eqref{pol}, by solving the boundary condition  \eqref{NBC}, 
the profile of  $Q$ is given by 
$X=-\lambda\eta$, where we have defined $\lambda=\frac{{\cal T}}{\s{1-{\cal{T}}^2}}$, and introduced new coordinates $U=T-X$ and $V=T+X$.  The metric on $Q$ is given by 
that of Poincare AdS$_2$ (see the left panel in Fig.~\ref{holsetupfig}),
\ba
ds^2=\frac{(1+\lambda^2)d\eta^2-dT^2}{\eta^2}.  
\label{eefq}
\ea
Thus, the gravity dual in terms of the $(U,V,\eta)$ coordinates is given by a part of Poincare AdS$_3$
defined by $X+\lambda \eta>0$.  Note that the surface $Q$ at 
the boundary $z=0$ coincides with the mirror trajectory $v=p(u)$ via the map \eqref{cordch}.

The gravity dual in terms of the coordinates $(u,v,z)$, which is given by the metric \eqref{metkads} and is sketched in the right panel of Fig.~\ref{holsetupfig}, only covers the region $U<0$ as $U=p(u)$ is always negative for any $u$. 
This is the bulk extension of the mentioned inner horizon region shown in the right panel in Fig.~\ref{mvsetupfig}.
In the coordinates $(u,v,z)$, the metric of the brane $Q$ reads
\begin{align}
ds^2\!\!  &=\!\!\frac{dz^2}{z^2}\!\!+\!\!\left(\!\frac{p''}{zp'}\!\!+\!\!\frac{2\lambda \s{p'}}{z^2}\!\right)\! dudz
+\!\!\left(\!\frac{p''^2}{4p'^2}\!\!-\!\!\frac{p'}{z^2}\!\!+\!\!\frac{\lambda p''}{z\s{p'}}\!\right)\! du^2. 
\label{qmet}
\end{align}
This covers only the part $T<-\lambda\eta$ of \eqref{eefq}.
In this way, the gravity dual of the moving mirror has a horizon, analogous to a single sided AdS black hole.

In general, the holographic entanglement entropy \cite{RT,HRT} in AdS/BCFT can be computed \cite{Ta} as 
\ba
S_A=\frac{1}{4G_N}\mbox{Min}_{\Gamma_A}\left[A(\Gamma_A)\right],  
\label{heefm}
\ea
where $A(\Gamma_A)$ is the length of $\Gamma_A$ which satisfies 
$\de \Gamma_A=\de A\cup \de I_s$, where $I_s$ (i.e. island) is a region on the surface $Q$.
The three dimensional Newton constant is denoted by $G_N$.
The minimum in \eqref{heefm} is taken over all possible choices of $I_s$ and $\Gamma_A$. 
When $A$ is an interval $[x_0,x_1]$ at time $t$, there appear to be two candidates for $\Gamma_A$. One is a connected 
geodesic between $x=x_0$ and $x=x_1$. The other one is a union of two disconnected ones, each of which departs from 
$x=x_0$ (or $x=x_1$) and ends on $Q$, respectively. 

\begin{figure}[h]
  \centering
  \includegraphics[width=0.2\textwidth]{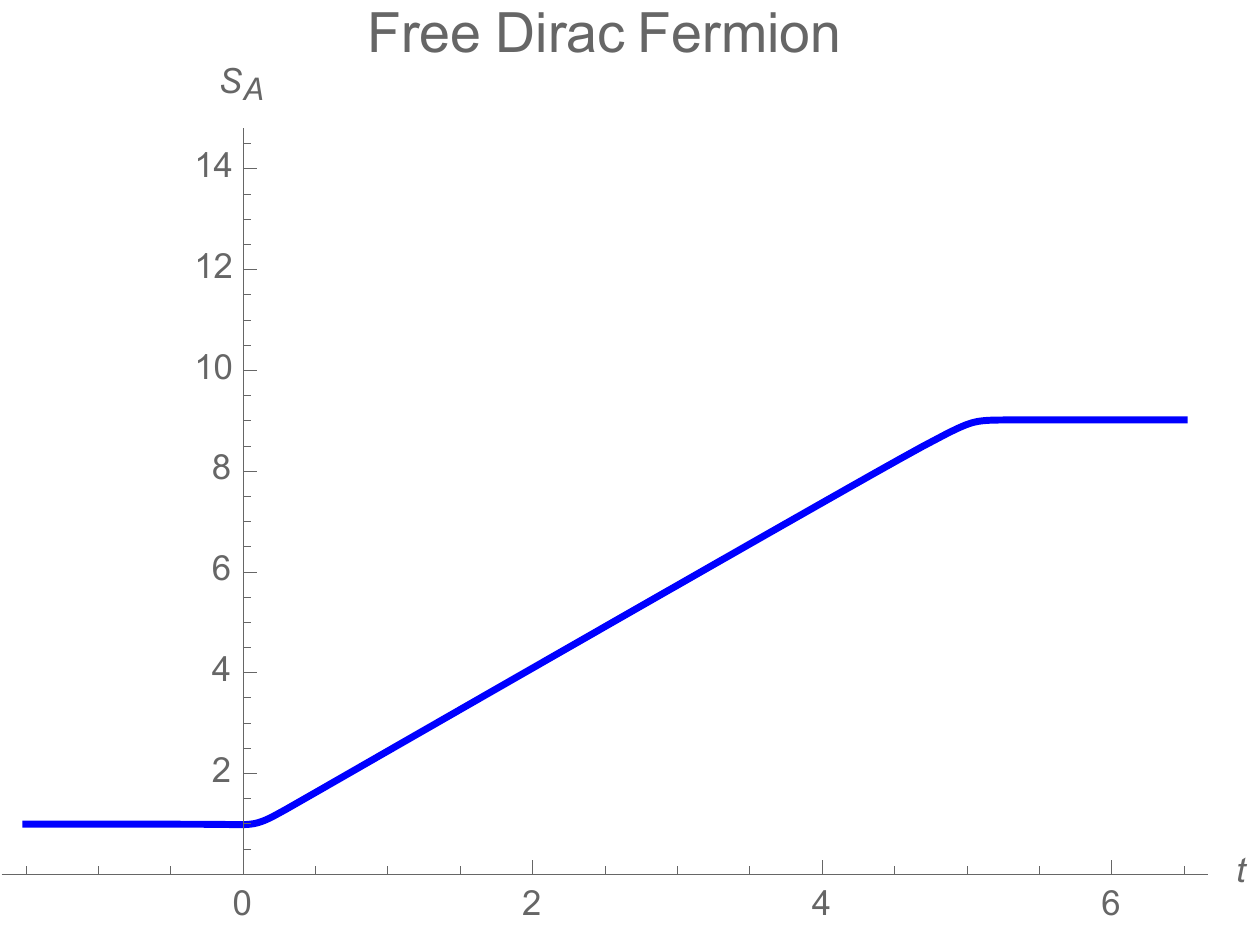}\qquad
  \includegraphics[width=0.2\textwidth]{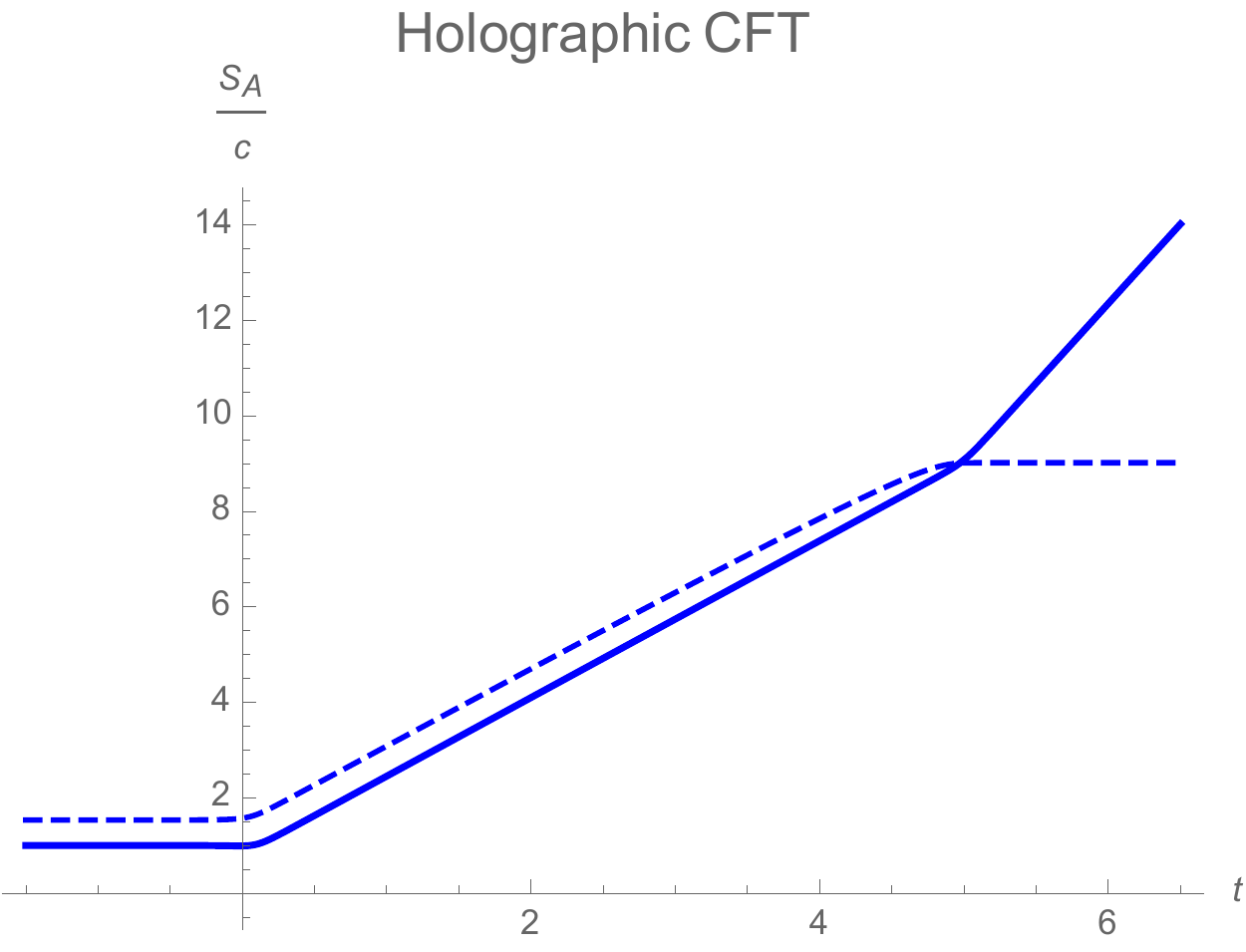}
  \caption{The time evolution of entanglement entropy for moving mirror \eqref{traj} in free Dirac fermion CFT (left) and in holographic CFT (right). Here, we set the subsystem to be $A=[Z(t)+0.1,Z(t)+10]$ with $\beta=0.1$, $\epsilon=0.1$ and $S_\text{bdy} = 0$. In the right, the thick line and the dashed line show the disconnected and connected entanglement entropy, respectively.}
\label{eefig}
\end{figure}

In our setup, they are explicitly given by 
$S_A=\mbox{Min}[S^\text{con}_A,S^\text{dis}_A]$, where the disconnected and connected geodesic contributions $S^\text{dis}_A$ and $S^\text{con}_A$ read
\begin{align}
S^\text{dis}_A &=
\frac{c}{6} \log\frac{t\!+\!x_0\!-\!p(t\!-\!x_0)}{\ep\s{p'(t\!-\!x_0)}}\! + 
\frac{c}{6} \log\frac{t\!+\!x_1\!-\! p(t\!-\!x_1)}{\ep\s{p'(t\!-\!x_1)}} \notag \\
&+ 2 S_\text{bdy}, \notag \\
S^\text{con}_A &=\frac{c}{6}\log\frac{(x_1-x_0)\left(p(t-x_0)-p(t-x_1)\right)}{\ep^2\s{p'(t-x_0)p'(t-x_1)}}.
\label{heecd}
\end{align}
The boundary entropy is a function of the tension and is given by $S_\text{bdy}=\frac{c}{6}\log\s{(1+{\cal T})/(1-{\cal T})}$. 
When $A$ is semi infinite, i.e. $x_1\to \infty$, we always have $S_A=S^\text{dis}_A$, and this reproduces \eqref{sahee}.
When $A$ is a finite interval, $S^\text{dis}_A$ is initially favored and this gives the linear growth as in \eqref{apphpee}. 
At later time, $S^\text{con}_A$ is favored and this leads to a saturation as depicted in the right panel of Fig.~\ref{eefig}.
We have also plotted $S_A$ for the massless free Dirac fermion case, which is shown in the left panel of Fig.~\ref{eefig}.
Refer to Appendix A for detailed computations of $S_A$ in the free Dirac fermion and holographic CFT case.

%%%%%%%%%%%%%%%%%%%%%%%%%%%%%
\section{Page curve from moving mirror}
%%%%%%%%%%%%%%%%%%%%%%%%%%%%%

\begin{figure}[h]
  \centering
  \includegraphics[width=0.15\textwidth]{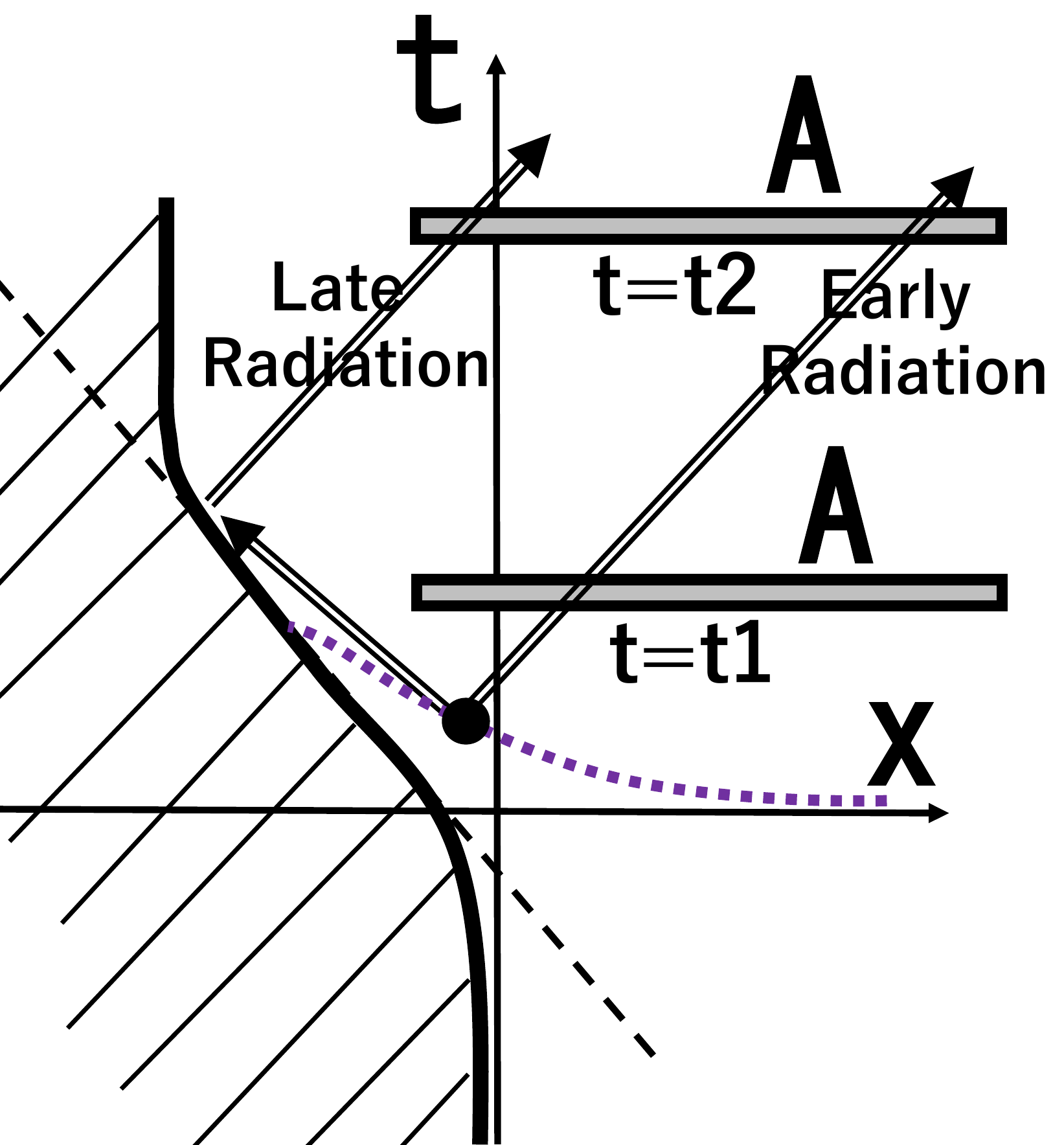}\qquad
  \includegraphics[width=0.22\textwidth]{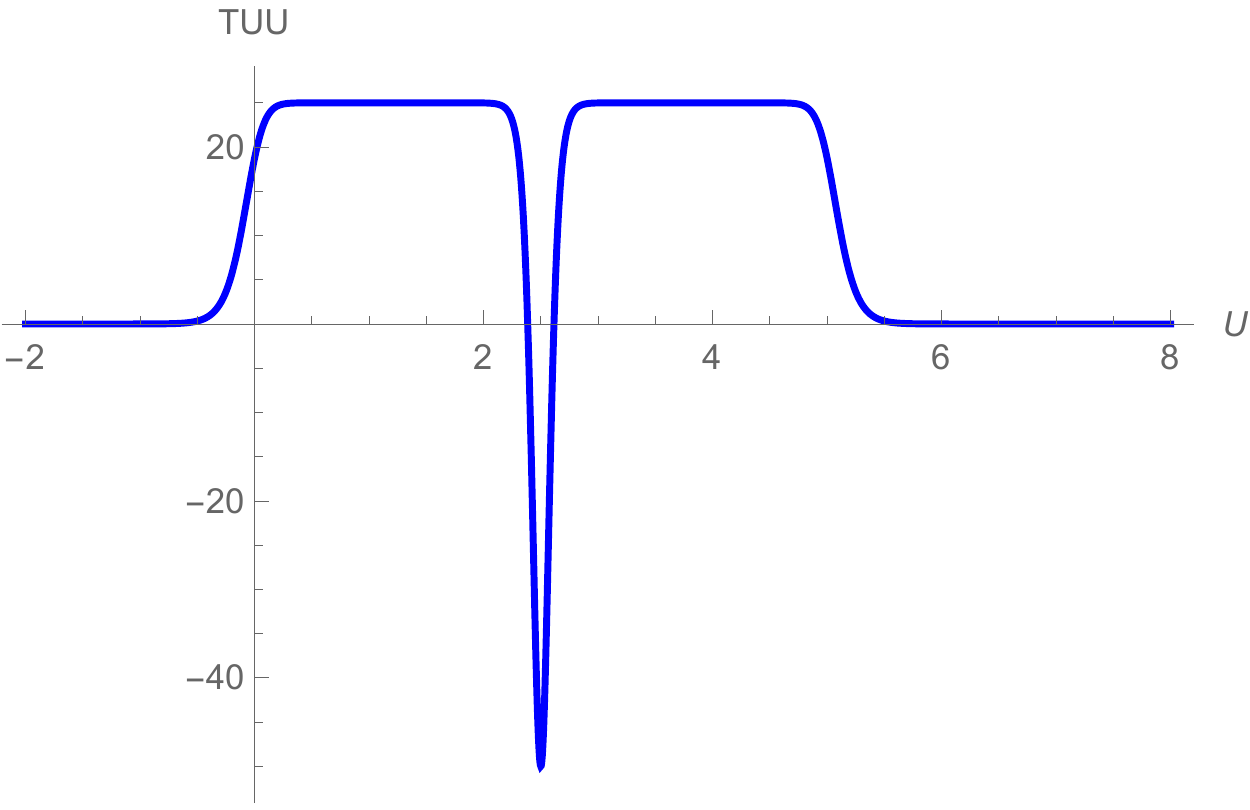}
  \caption{The profile of moving mirror, with the creation of entangled pairs and their reflection at the mirror, is 
depicted in the left picture. The right graph is the energy density $T_{uu}$ plotted as a function of $u$.
We set $\beta=0.1$ and $u_0=5$.}
\label{pagefig}
\end{figure}

A typical moving mirror model which mimics an evaporating black hole is found by setting
\ba
p(u)=-\beta\log(1+e^{-\frac{u}{\beta}})+\beta\log(1+e^{\frac{u-u_0}{\beta}}),  \label{kexfr}
\ea
whose mirror trajectory $x=Z(t)$ and energy flux $T_{uu}$ are depicted in Fig.~\ref{pagefig}.
When $\beta$ is small, we can approximate the trajectory as
$Z(t)\simeq 0$ for $t<0$,  $Z(t)\simeq -t$ for $0<t<u_0/2$, and
$Z(t)=-u_0/2$ for $t>u_0/2$. The energy flux is nonvanishing, $T_{uu}\simeq \frac{c}{48\pi\beta^2}$, namely, only for the period $0<u<u_0$.

\begin{figure}[h]
  \centering
  \includegraphics[width=0.2\textwidth]{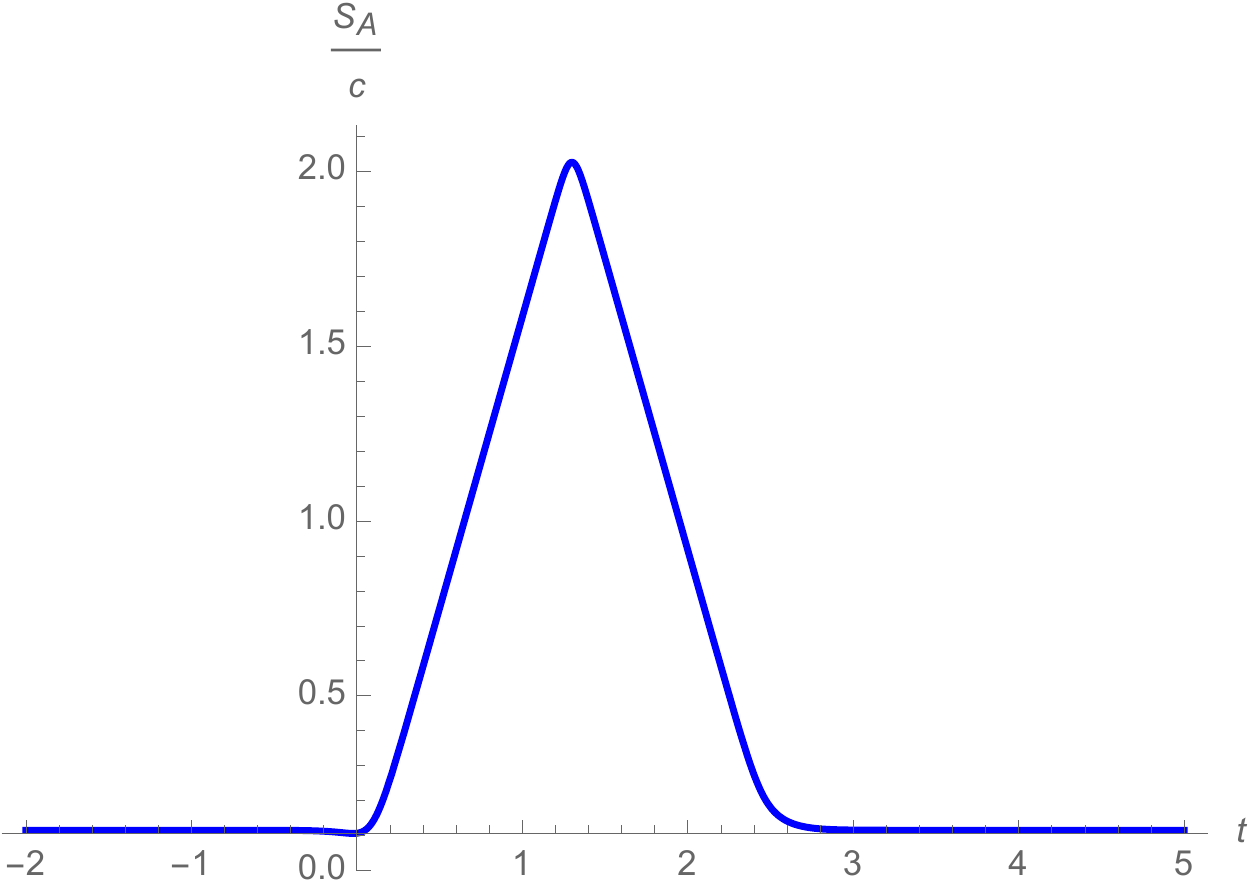}\qquad
  \includegraphics[width=0.2\textwidth]{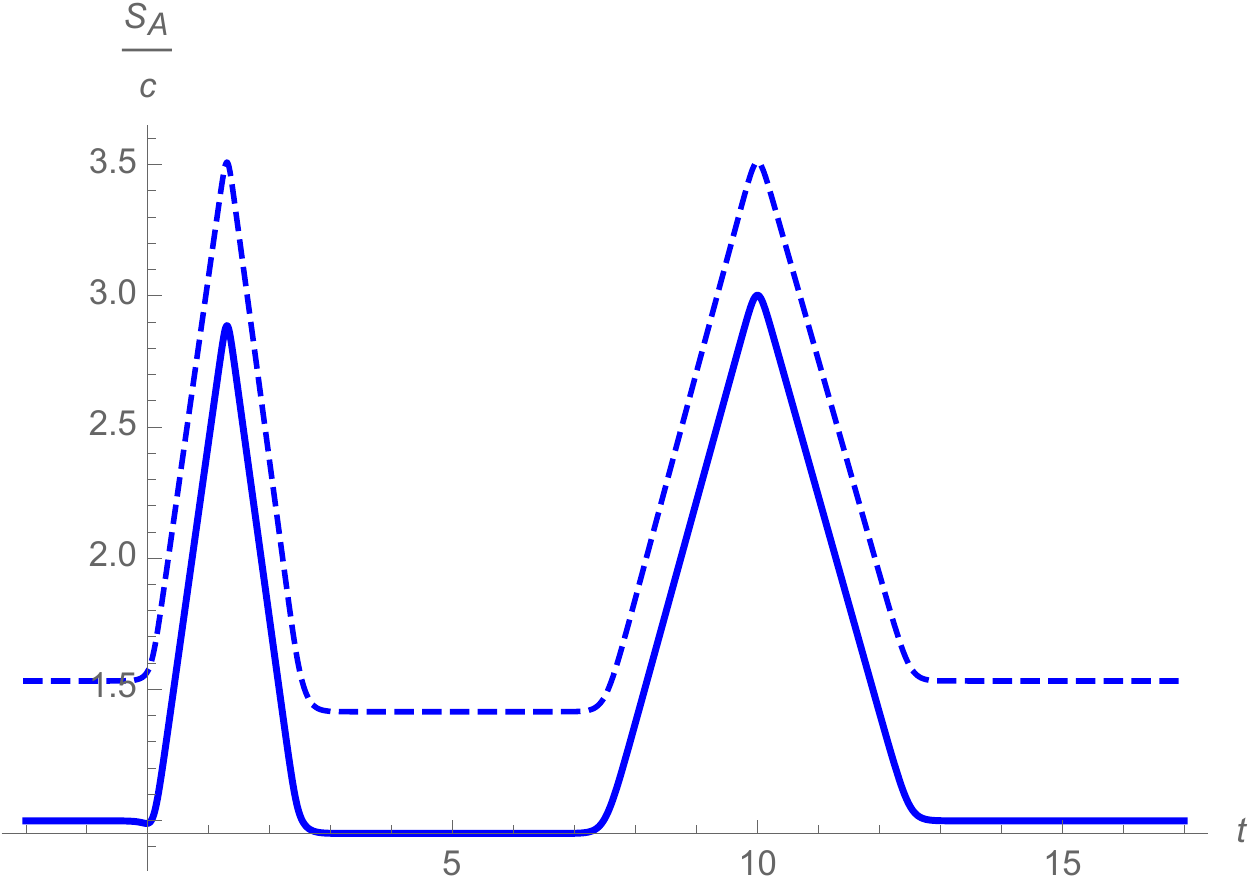}
  \caption{The time evolution of holographic entanglement entropy follows the Page curve. We choose $A$ to be a semi infinite line $A=[Z(t)+0.1,\infty]$ in the left and a finite interval $A=[Z(t)+0.1,Z(t)+10]$
in the right. The thick and dashed curves describe $S^\text{dis}_A$ and $S^\text{con}_A$, respectively. 
We set $\beta=0.1, u_0=5$, $\ep=0.1$ and $S_\text{bdy} = 0$.}
\label{pagetfig}
\end{figure}

We can again calculate the holographic entanglement entropy $S_A=\mbox{Min}[S^\text{con}_A,S^\text{dis}_A]$, using \eqref{heecd} as plotted in Fig.~\ref{pagetfig}. In particular, when $A$ is a semi infinite line, $S_A$ takes the form of the Page curve. For $0<t<u_0/4$, $S_A$ grows linearly $\frac{dS_A}{dt}\simeq \frac{c}{6\beta}$, and for $u_0/4<t<u_0/2$, it decreases
linearly, i.e. $\frac{dS_A}{dt}\simeq -\frac{c}{6\beta}$. Note that as is clear from Fig.~\ref{pagetfig}, the disconnected result
$S^\text{dis}$ gives the dominant contribution (refer to \cite{Hotta:2015huj} for an earlier calculation of the connected result 
$S^\text{con}$).

The initial linear growth of $S_A$ can be understood as in the previous example 
\eqref{apphpee} by considering entangled pair production along the curve $v+p(u)=0$. Moreover, the linear decay of $S_A$ is 
explained by reflections of the left moving partner, as shown in the left picture of Fig.~\ref{pagefig}.
When $A$ is a finite interval, we have two Page peaks. The first peak occurs when only the originally right moving particles are crossing $A$. The second peak appears when only the reflected particles are crossing $A$.

%%%%%%%%%%%%%%%%%%%%%%%%%%%%%
\section{Brane world gravity and island}
%%%%%%%%%%%%%%%%%%%%%%%%%%%%%

\begin{figure}
  \centering
\includegraphics[width=0.19\textwidth]{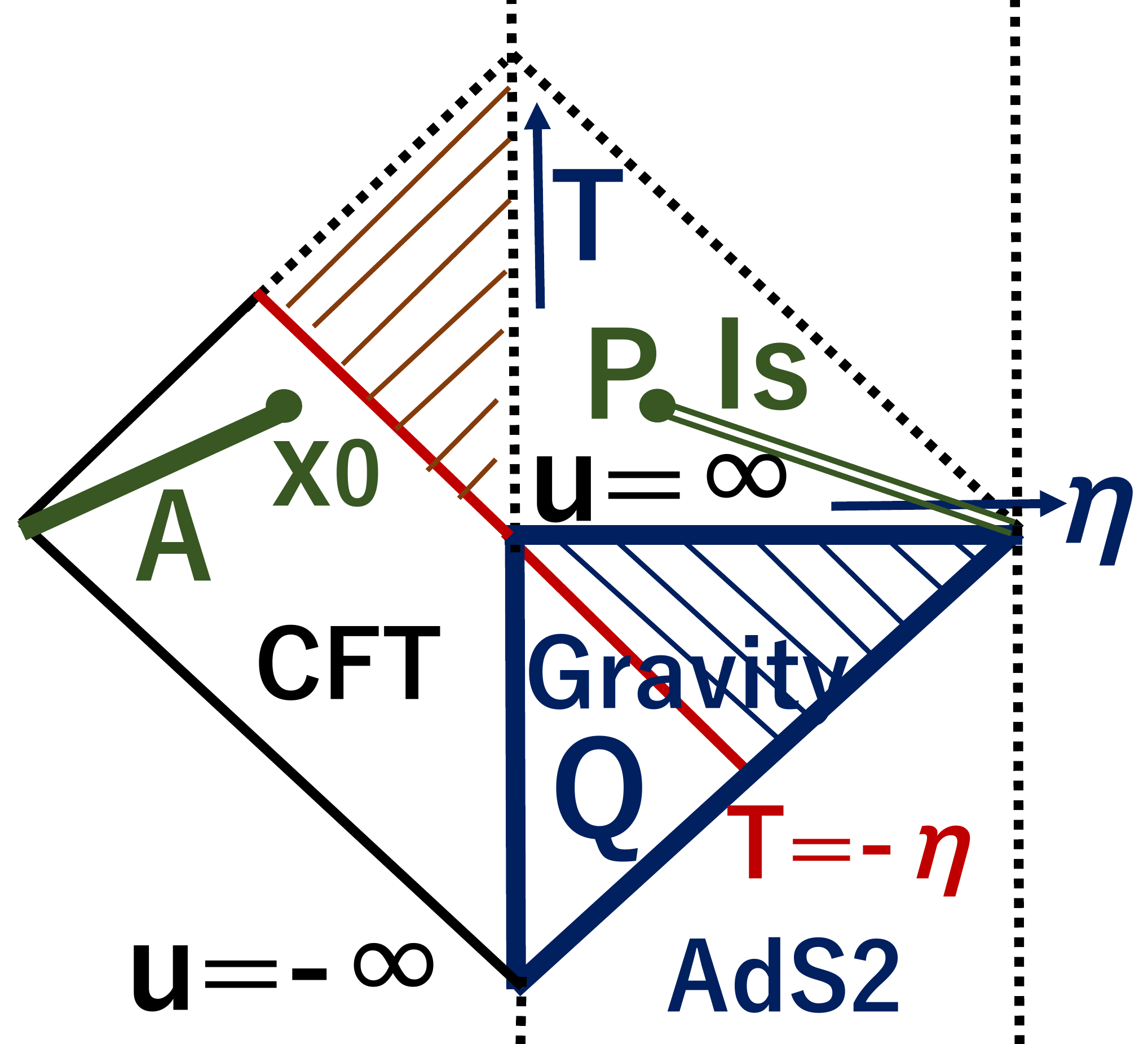}\qquad
  \includegraphics[width=0.17\textwidth]{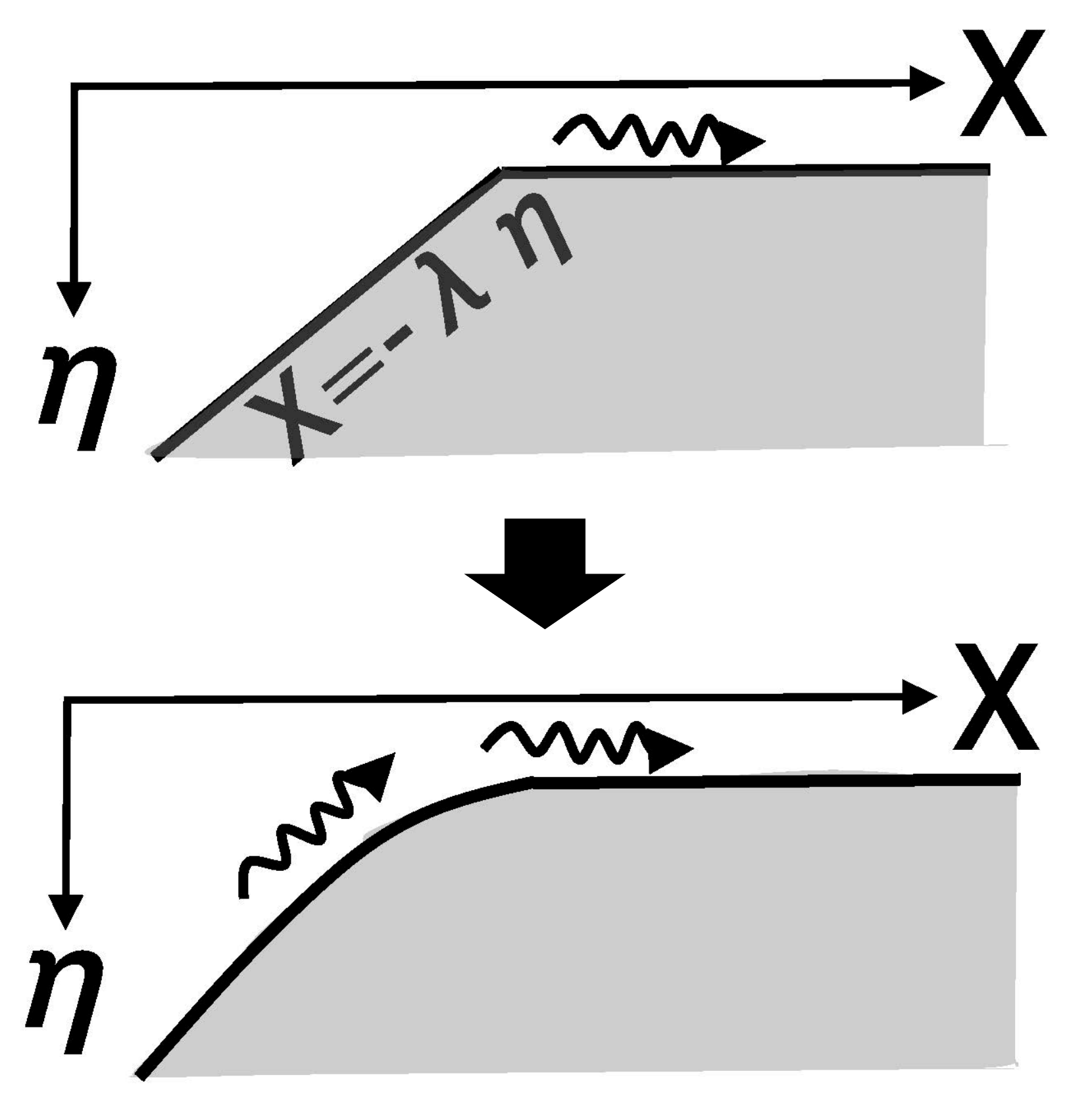}
  \caption{The left picture shows the global spacetime of a CFT (left triangle) and the brane world gravity on $Q$ (right triangle), which are attached along the mirror trajectory. The island (doubled green line) is shown as well. The right picture sketches the deformation of the gravity dual of the moving mirror (top) into that with black hole radiation (bottom).}
\label{eeafig}
\end{figure}

The gravity dual of our moving mirror setup can be interpreted in an alternative way by regarding the surface $Q$ as an end of the world brane in the brane world setup \cite{Randall:1999ee,Randall:1999vf,Karch:2000ct}. This situation is depicted in the left panel of Fig.~\ref{eeafig}. According to this interpretation, the CFT defined in the region $x\geq Z(t)$ will be coupled to a two dimensional gravity theory on $Q$. By estimating the effective Newton constant on $Q$ via Kaluza-Klein reduction \cite{Randall:1999ee,Randall:1999vf,Karch:2000ct,Akal:2020wfl}, which we denote by $G^{(Q)}_N$, we can show
\ba
\frac{1}{4G^{(Q)}_N}=S_\text{bdy}.  
\label{bdye}
\ea
The gravitational entropy of AdS$_2$, i.e. the brane $Q$, will thus be equal to the boundary entropy $S_\text{bdy}$.

We can regard $S_A$ in \eqref{sahee} as the entanglement entropy of the subregion $A$ in a system consisting of a CFT on $x\geq Z(t)$ and a gravitational theory on $Q$, glued along the moving mirror. Then, we can interpret the first and second term in \eqref{sahee} as the bulk entropy contribution $S^\text{bulk}_{A\cup I_s}$ and the area term $\frac{\text{Area}(\de I_s)}{4G_N}$, respectively, in the island formula \cite{Penington:2019npb,Almheiri:2019psf,Almheiri:2019hni,Penington:2019kki,Almheiri:2019qdq}. Note that here we use the standard formula for computing holographic entanglement entropy without invoking the quantum extremal surface prescription. The density matrix under consideration is pure and radiation is manifestly unitary.

In our moving mirror model, the entropy \eqref{sahee} shows linear growth
in the region defined by the equation $v+p(u)>0$. This part in the entanglement entropy would arise from the island. This region is not covered by the coordinate patch \eqref{qmet}, as sketched in the left panel of Fig.~\ref{eeafig}. 

An interesting feature in our setup is the entanglement between the gravitational 
theory on $Q$ and the CFT by the amount \eqref{bdye}. Also it is important to 
note that the presence of energy flux from the boundary is different from standard 
BCFTs (Cardy states \cite{CardyS}) which have no energy 
flux condition $T(w)-\bar{T}(\bar{w}) = 0$ at the boundary.

The radiation in the CFT looks similar to the setups \cite{Penington:2019npb,Almheiri:2019psf,Almheiri:2019hni,Penington:2019kki,Almheiri:2019qdq}, where the Page curve was derived.
However, unlike these, in our BCFT model, we find that there is no radiation present in the gravitational system on $Q$. 
That is, the flux does not come from the gravitational system, but is created on the boundary. Indeed, the holographic energy 
stress tensor \cite{Balasubramanian:1999re} on $Q$ is proportional to $h_{ab}$ as it follows from \eqref{NBC}. 
Hence, it can just be regarded as a negative cosmological constant. 
The absence of radiation from $Q$ is obviously consistent 
with the fact that the mirror is completely reflective. 
We can regard our analysis as a derivation of the Page curve in the strong coupling regime of gravity, while \cite{Penington:2019npb,Almheiri:2019psf,Almheiri:2019hni,Penington:2019kki,Almheiri:2019qdq} focus on the weak coupling regime.

Indeed, by changing the profile of the brane $Q$, we can deform our setup of the moving mirror such that it incorporates radiation resulting from the gravitational sector on $Q$, see right panel of Fig.~\ref{eeafig}. If we modify the surface $Q$ to make it close to the standard AdS boundary located at $z=\ep$, the matter energy stress tensor 
on $Q$ will be approximated by \eqref{EMTm}. This provides a special and concrete example of the setup considered in \cite{Almheiri:2019hni}, see also \cite{Almheiri:2019yqk,Rozali:2019day,Chen:2019uhq,Chen:2020uac,Bousso:2020kmy,Akal:2020wfl,Chen:2020tes} for related works.
The holographic dual of this modified setup is a two dimensional CFT coupled to two dimensional gravity.
One advantage of our procedure is that our calculation based on the BCFT analysis is much easier than the one based on the conformal welding problem \cite{Almheiri:2019psf,Almheiri:2019qdq}. An interesting future direction will be relating the two realizations in an explicit way \cite{Future}.

%%%%%%%%%%%%%%%%%%%%%%%%%%%%%
\section{Conclusion}
%%%%%%%%%%%%%%%%%%%%%%%%%%%%%

In this article, we have presented a gravity dual of two dimensional CFT with a moving mirror, which mimics black hole
formation and evaporation. We have explicitly calculated the time evolution of entanglement entropy in the presence of the mirror. We have found that it follows the ideal Page curve. This can be explained by the creation of entangled particles, their propagation, and reflection from the mirror.  We have also discussed that modifying the profile of the end of the world brane in the gravity dual results in a model for two dimensional black hole radiation. In order to understand unitary evolution for realistic black hole evaporation, we will have to incorporate the singularity. We expect that the presence of spacelike boundaries in the CFT and its gravity dual \cite{Akal:2020wfl,Chen:2020tes} will be relevant \cite{Future}.

%\vspace{5mm}
%%%%%%%%%%%%%%%%%%%%%%%%%%%%%
{\bf Acknowledgements} 
%%%%%%%%%%%%%%%%%%%%%%%%%%%%%
We are grateful to Tatsuma Nishioka, Kotaro Tamaoka and Tomonori Ugajin 
for useful comments on a draft of this paper.
IA, YK, and ZW are supported by the Japan Society for the Promotion of Science (JSPS).
IA is supportted by the Alexander von Humboldt (AvH) foundation.
IA and TT are supported by Grant-in-Aid for JSPS Fellows No.~19F19813.
YK is supported by Grant-in-Aid for JSPS Fellows No.~18J22495.
NS is also supported by JSPS KAKENHI Grant No.~JP19K14721.
TT is supported by the Simons Foundation through the ``It from Qubit'' collaboration.  
TT is supported by Inamori Research Institute for Science and 
World Premier International Research Center Initiative (WPI Initiative) 
from the Japan Ministry of Education, Culture, Sports, Science and Technology (MEXT). 
SN and TT are supported by JSPS Grant-in-Aid for Scientific Research (A) No.~16H02182. 
TT is also supported by JSPS Grant-in-Aid for Challenging Research (Exploratory) 18K18766. 
ZW is supported by the ANRI Fellowship and Grant-in-Aid for JSPS Fellows No.~20J23116.

%%%%%%%%%%%%%%%%%%%%%%%%%%%%%%%%%%%

%%%%%%%%%%%%%%%%%%%%%%%%%%%%%
\appendix
%%%%%%%%%%%%%%%%%%%%%%%%%%%%%

\section{Appendix A: Calculating entanglement entropy in two dimensional CFTs 
with a moving mirror}

In the following, we present details of our computations of entanglement entropy in the presence of a moving mirror for CFTs
in two dimensions. The replica trick is often used to compute the entanglement entropy in quantum field theories \cite{CC04}. For two dimensional CFTs, 
in particular, the $n$-th R\'{e}nyi entropy of a single interval, $A=[x_0,x_1]$, can be evaluated as 
\ba
S^{(n)}_A = \frac{1}{1-n}\log \langle\sigma_{n}(t,x_0)\bar{\sigma}_{n}(t,x_1)\rangle.
\label{eq:renyi}
\ea
Here, $\sigma_{n}(t,x_0)$ and $\bar{\sigma}_{n}(t,x_1)$ are twist operators which act as primaries with conformal weights 
\ba
h_n=\bar{h}_n = \frac{c}{24} \left(n-\frac{1}{n}\right).
\ea
The expectation value $\langle\cdots\rangle$ is evaluated by the path integral on the manifold with a moving mirror. Once an analytical form of the R\'{e}nyi entropy is found, we can perform the analytic continuation to obtain the entanglement entropy. 
To evaluate the 2-point function of twist operators, we use the following conformal map
\begin{align}
    \tilde{u} = p(u), \quad \tilde{v} = v
\end{align}
to map the moving mirror setup into the right half plane (RHP) of $\mathbb{R}^{1,1}$, which is parameterized by 
\begin{align}
    (\tilde{t}, \tilde{x}) = \left(\frac{\tilde{u}+\tilde{v}}{2},\frac{-\tilde{u}+\tilde{v}}{2}\right).
\end{align}
Under this conformal transformation, we have
\begin{align}
    &\langle\sigma_{n}(t,x_0)\bar{\sigma}_{n}(t,x_1)\rangle 
    =\\
    &\left(p'(u_0)p'(u_1)\right)^{h_n} \cdot
 \langle\tilde{\sigma}_{n}\left(\tilde{t}_0,\tilde{x}_0\right)\tilde{\bar{\sigma}}_{n}\left(\tilde{t}_1,\tilde{x}_1\right)\rangle_{\rm RHP}. \notag 
    \label{eq:2pf}
\end{align}
Thus, our task is reduced to evaluating a 2-point function of twist operators on the RHP.
As concrete examples, in what follows, we will do so for free massless Dirac fermion CFT and holographic CFT.

\subsection{Dirac fermion CFT}
For the massless Dirac fermion CFT, after performing doubling trick and bosonization (see e.g. \cite{Casini:2009sr}), we have
\begin{align}
    &\langle\tilde{\sigma}_{n}\left(\tilde{t}_0,\tilde{x}_0\right)\tilde{\bar{\sigma}}_{n}\left(\tilde{t}_1,\tilde{x}_1\right)\rangle_{\rm RHP} \nonumber \\
    =& \sqrt{ \langle
    \tilde{{\sigma}}_{n}\left(\tilde{t}_1,-\tilde{x}_1\right)
    \tilde{\bar{\sigma}}_{n}\left(\tilde{t}_0,-\tilde{x}_0\right)
    \tilde{\sigma}_{n}\left(\tilde{t}_0,\tilde{x}_0\right)  \tilde{\bar{\sigma}}_{n}\left(\tilde{t}_1,\tilde{x}_1\right) \rangle_{\mathbb{R}^{1,1}}}
      \nonumber\\
    \propto& \left( \frac{(\tilde{u}_0-\tilde{v}_1)(\tilde{v}_0-\tilde{u}_1)}{(\tilde{u}_0-\tilde{u}_1)(\tilde{v}_0-\tilde{v}_1)(\tilde{u}_0-\tilde{v}_0)(\tilde{u}_1-\tilde{v}_1)} \right)^{2h_n}.
\end{align}
By plugging this into \eqref{eq:2pf} and \eqref{eq:renyi} with proper $p(u)$, and taking the limit $n\rightarrow 1$, we will get the entanglement entropy for the moving mirror setup in Dirac fermion CFT. 

\subsection{Holographic CFT}
In holographic CFT, the 2-point function can be evaluated as (see e.g. \cite{Sully:2020pza})
\begin{align}
    &\langle\tilde{\sigma}_{n}\left(\tilde{t}_0,\tilde{x}_0\right)\tilde{\bar{\sigma}}_{n}\left(\tilde{t}_1,\tilde{x}_1\right)\rangle_{\rm RHP} \\
    =& \max
  \begin{cases}
    \langle\tilde{\sigma}_{n}\left(\tilde{t}_0,\tilde{x}_0\right)\tilde{\bar{\sigma}}_{n}\left(\tilde{t}_1,\tilde{x}_1\right)\rangle_{\mathbb{R}^{1,1}} \\
    g^{2(1-n)} \prod_{i\in\{0,1\}}
    \langle\tilde{\sigma}_{n}\left(\tilde{t}_i,\tilde{x}_i\right)\tilde{\bar{\sigma}}_{n}\left(\tilde{t}_i,-\tilde{x}_i\right)\rangle_{\mathbb{R}^{1,1}}^{1/2}
    \label{corbc}
  \end{cases}. \notag
\end{align}
Here, $g$ is a constant which depends on the details of the boundary condition on the moving mirror and results in the boundary entropy $S_\text{bdy}=\log g$. 
These two cases correspond to the connected channel and the disconnected channel, respectively. We can obtain \eqref{corbc} by assuming the simple Wick contraction rule 
(so called generalized free field prescription) by factoring the correlation functions into two point functions of the twist operators. This is justified due to the large 
$c$ factorization property of holographic CFTs. 
The disconnected channel arises by considering the mirror operator across the boundary. Therefore, we call the von Neumann entropy obtained from both channels the connected entanglement entropy, $S_A^\text{con}$, and the disconnected entanglement entropy, $S_A^\text{dis}$, respectively. The physical entanglement entropy is
\begin{align}
	S_{A} = \mathrm{Min} \left[ S_A^\text{con}, S_A^\text{dis} \right].
\end{align}
Remembering that the 2-point function on $\mathbb{R}^{1,1}$ is given by
\begin{align}
    \langle\tilde{\sigma}_{n}\left(\tilde{t},\tilde{x}\right)\tilde{\bar{\sigma}}_{n}\left(\tilde{t}',\tilde{x}'\right)\rangle_{\mathbb{R}^{1,1}} = \frac{1}{|(\tilde{t}-\tilde{t}')^2-(\tilde{x}-\tilde{x}')^2|^{2h_n}},
\end{align}
plugging these into \eqref{eq:2pf} and \eqref{eq:renyi} with proper $p(u)$ and taking the limit $n\rightarrow 1$, we get the connected and the disconnected entanglement entropy for the moving mirror setup in holographic CFT. 
This leads to the results in \eqref{heecd}. As discussed in the main text, for deriving \eqref{heecd} we have performed a holographic calculation by evaluating the length of geodesics in the bulk. The computation here is performed on the CFT side. 
In fact, we can easily find out that evaluating the 2-point function in the (dis)connected channel on the CFT side is exactly equivalent to evaluating the length of the (dis)connected geodesic on the AdS side.

\end{document}